\documentclass[onecolumn,preprint]{aastex631}
\usepackage{chemformula}
\let\ce\ch
\usepackage[caption=false]{subfig}
\usepackage{hyperref}
\usepackage{xcolor}
\definecolor{medium-blue}{rgb}{0,0,1}
\hypersetup{colorlinks, urlcolor={medium-blue}}
\urldef{\footurl}\url{https://news.arizona.edu/story/ua-researcher-captures-rare-radar-images-comet-46pwirtanen}

\accepted{April 10, 2023}

\newcommand{\edits}[1]{\textcolor{black}{#1}}
\newcommand{\secondedits}[1]{\textcolor{black}{#1}}
\newcommand{\thirdedits}[1]{\textcolor{black}{#1}}

\shorttitle{46P and Icy Grains}
\shortauthors{Kareta and Noonan et al.}
\begin{document}
\title{Ice, Ice, Maybe? Investigating 46P/Wirtanen's Inner Coma For Icy Grains}

\correspondingauthor{Theodore Kareta}
\email{tkareta@lowell.edu}

\author{Theodore Kareta}
\altaffiliation{These authors contributed equally to this work.}
\affiliation{Lowell Observatory, Flagstaff, Arizona 86001, USA}

\author{John W. Noonan}
\altaffiliation{These authors contributed equally to this work.}
\affiliation{Department of Physics, Auburn University, Auburn, Alabama, 36849, USA}

\author{Walter M. Harris}
\affiliation{Lunar and Planetary Laboratory, University of Arizona, Tucson, Arizona 85721-0092, USA}

\author{Alessondra Springmann}
\affiliation{Southwest Research Institute,
Boulder, Colorado, 80302, USA}

\begin{abstract}
The release of volatiles from comets is usually from direct sublimation of ices on the nucleus, but for very or hyper-active comets other sources have to be considered to account for the total production rates. In this work, we present new near-infrared imaging and spectroscopic observations of 46P/Wirtanen taken during its close approach to the Earth on 2018 December 19 with the MMIRS instrument at the MMT Observatory to search for signatures of icy or ice-rich grains in its inner coma that might explain its previously reported excess water production. The morphology of the images does not suggest any change in grain properties within the field of view, and the NIR spectra do not show the characteristic absorption features of water ice. Using a new MCMC-based implementation of the spectral modeling approach of \citet{protopapa_icy_2018}, we estimate the areal water ice fraction of the coma to be $<0.6\%$. When combined with slit-corrected Af$\rho$ values for the J, H, and K bands and previously measured dust velocities for this comet, we estimate an icy grain production rate of less than \edits{4.6} kg s$^{-1}$. This places a strict constraint on the water production rate from \thirdedits{pure} icy grains in the coma, and in turn we find that for the 2018-2019 apparition approximately 64\% of 46P's surface was actively sublimating water near perihelion. We then discuss 46P's modern properties within the context of other (formerly) hyper-active comets to understand how these complex objects evolve.
\end{abstract}

\section{Introduction}\label{Intro}
Comets are the most easily studied icy \secondedits{planetesimals}, and their modern properties are thought to encode information about their formation in the outer protosolar disk billions of years ago. The challenge is thus differentiating between which parts of their modern state -- activity level, coma chemistry, nucleus size and topography, etc. -- are primordial and which are the result of the subsequent evolution. While some comets are 'hyperactive' (meaning that they produce more water vapor than sublimation of an equivalent surface area of pure \edits{water} ice would, like 103P/Hartley 2 \citep{a2011epoxi,protopapa2014water}, \edits{thus requiring an active fraction of over 100$\%$}), others are nearly or totally inactive like 249P/Linear or 2003 EH$_1$ \citep{2021KaretaEH1,knight_multi-spacecraft_2020}. It is thus of great interest to study comets which appear to be changing activity states in the present to understand how these processes might apply to the population of comets in general. A great example and the subject of this paper is 46P/Wirtanen, the original target of the European Space Agency's \textit{Rosetta} mission, whose overall production of volatile species has been dropping consistently since at least the 1990s \citep{knight2021narrowband}. A similar water production rate drop was measured \edits{by} \citet{combi_comet_2020}. 46P was very likely a 'hyperactive' comet \citep{farnham1998narrowband} \edits{thanks to} its relatively small nucleus (\edits{r}$\sim0.6$km, \cite{lamy1998nucleus,lis_terrestrial_2019}) in the 1990s and is still a rather high-activity comet today, even considering the large overall production rate drop reported in \citet{knight2021narrowband} and \citet{combi_comet_2020}. \edits{\citet{groussin2003activity} showed that 46P had an active fraction between 70 and 100\% within 1.5 au during its 1996-1997 apparition, sustaining 100\% throughout it's perihelion passage. For the recent 2018-2019 apparition \citet{2021PSJ.....2..176P} implemented the sublimation model of \citet{a1984comet} to interpret the water production rate of \citet{knight2021narrowband} and found that 46P's active fraction had dropped to 58\%.} If we could understand what is driving this modern rapid evolution of 46P, it might present an interesting point of comparison between the hyperactive comets, the more conventional comets, and low-activity or dormant comets. Our ability to monitor the evolving production rates of 46P offers insight into the processes that drive hyperactivity and how they change with time.

Some comets show evidence for solid grains of pure or nearly-pure water ice in their comae. These grains have been detected in-situ at 9P/Tempel 1 in the ejecta from the \textit{Deep Impact} experiment \citep{a2005deep, sunshine2007distribution} as well as in the inner coma of the aforementioned 103P/Hartley 2 during the \textit{Deep Impact eXtended Investigation (DIXI)} \citep{protopapa2014water}, in long period comets like C/2013 US10 (Catalina) \citep{protopapa_icy_2018} or C/2011 L4 (PanSTARRS) \citep{yang_multi-wavelength_2014}, and even in the coma of the active Centaur P/2019 LD2 (ATLAS)\citep{2021KaretaLD2}. \edits{\citet{protopapa_icy_2018} compared the results from these studies and} generally concluded that icy grains, especially those near $\sim1\mu{m}$ in size, are likely a common component of most or all cometary nuclei. The sublimation of icy grains is a natural source for the `excess' water production inferred for hyperactive comets, but direct observations of those grains is challenging for the same reason -- \citet{protopapa_icy_2018} estimates that \thirdedits{pure water ice grains} \secondedits{on the order of 1$\mu$m in diameter have lifetimes on the order of 3$\times$10$^{4}$ -6$\times$10$^{11}$ s at 1.3–5.8 au,} \edits{heavily dependent on heliocentric distance, water ice particle size, and how impure the grains are.} Those detections at small heliocentric distances at 9P and 103P \edits{benefited from} the high spatial resolution afforded by close fly-bys of spacecraft, \edits{but $\mu$m-sized pure icy grains should still be detectable} at typical geocentric distances from the ground \edits{\citep{protopapa_icy_2018,2021PSJ.....2..176P}}. 

 Icy grains should begin sublimating significantly inside of where water sublimation begins to dominate cometary activity ($\sim$4 AU), and \citealt{protopapa_icy_2018} employ a sublimation model to suggest that beyond this distance these icy grains may be quite long lived. This suggestion is bolstered by the likely discovery of water ice in the coma of the active JFC Gateway Centaur P/2019 LD2 (ATLAS) by \citet{2021KaretaLD2}, as those authors interpreted the material being lofted was from the surface and near-surface due to the stable nature of its activity. As that object has very likely never visited the inner solar system \citep{2020RNAAS...4...74K, hsieh_potential_2020,steckloff2020p} and thus never experienced temperatures high enough to sublimate water significantly, it seems likely that the water ice grains really are a common component of the nuclei of comets; it simply becomes harder to detect as the grains sublimate faster and faster as the comet approaches the Sun. While the detections of ice inside the outburst and impact driven comae of 17P/Holmes \citep{yang_comet_2009} and 9P/Tempel 1 \citep{sunshine2007distribution} can be explained as icier material from the interior of the comets being excavated, the hyperactive comet 103P/Hartley 2 \edits{produced pure water ice grains during sustained activity} \citep{protopapa2014water}. The sublimation of icy grains in its coma (an extended source) provide a natural explanation for its large water production rate for its surface area. \edits{If that's the case, why aren't all comets ejecting icy grains if they really are a common material in cometary nuclei? Or is it the case that we simply don't have sufficient reflectance observations of comets at large heliocentric distances to identify the water ice grains?}

\secondedits{The close approach of comet 46P/Wirtanen in 2018/2019 provided an excellent opportunity to observe the inner coma of a comet from Earth-based observatories. One goal of the close approach was to search for icy grains like those found in the coma of its fellow high-activity comet 103P, taking advantage of the low perigee of just 0.07 AU to probe the inner coma. Grains like those found around 103P (d$\approx$1 micron and pure water ice) at 46P are expected to be detectable based on their lifetimes and observing conditions (as detailed in \cite{2021PSJ.....2..176P}).} The narrow-band imaging presented \secondedits{by} \citet{knight2021narrowband} was interpreted to show evidence for an extended source for OH and NH molecules, which those authors argued was best explained as being sourced by an icy grain population. Understanding the role that small icy grains currently play in 46P's inner coma is important for properly determining the current fractional active area of the nucleus, placing constraints on extended sources for potentially clathrated molecules that could be released from a sublimating icy grain population, and ultimately understanding how the life cycle of a hyperactive comet may end.  

In this paper we present near-infrared observations of 46P/Wirtanen taken just after close approach on December 19, 2018 with the MMIRS NIR multi-object slit spectrograph instrument at the 6.5-m MMT Observatory on Mount Hopkins in Arizona. We compare our results to those of \citet{2021PSJ.....2..176P}, \edits{who also used near-infrared spectra to search for ice, throughout the text. In short, those authors set stringent upper limits on the presence of water ice grains in the coma of 46P/Wirtanen based on six dates of near-infrared spectroscopy, including a non-detection of the very strong water ice band at $3.0\mu{m}$ \citep{2021PSJ.....2..176P}. Those authors argue that when combined with state-of-the-art \secondedits{spectroscopic and} grain lifetime models that pure (e.g. each grain is $100\%$ ice) water ice grains are incompatible with the data, but \secondedits{grains on the order of 1 $\mu$m in size} with $\sim0.5\%$ impurities or \secondedits{large chunks containing significant amounts of water ice at depth } compatible with their observations.} In Section \ref{Obs} we discuss the observational planning and geometry for the relevant days, the instrument settings, our reduction methods, and present an observing log \secondedits{as well as our reduced spectrum and spectral slope}. In Section \ref{Modeling}, we describe and apply a Markov-Chain Monte Carlo approach to spectral modeling and use it to calculate a realistic upper limit on the water ice areal fraction in the coma. \edits{Section \ref{Results} contains our derived spectral properties, upper limits for areal water ice fraction, and estimates for ice production rates.}  In Section \ref{Disc} we review the implications of the lack of water ice absorption in the extreme inner coma of 46P and the relationship to water production rates and volatile species morphologies. \edits{Finally, in Section \ref{Summary} we summarize our findings and detail future efforts. }

\section{Observations}\label{Obs}
\subsection{Observation Log and Data Reduction}
Observations were acquired on the early morning of December 19, 2018 with the MMIRS NIR instrument on the MMT telescope \citep[]{mcleod_mmt_2012}. \edits{To search for the water ice absorption features at $\sim1.5$ and $\sim2.0\mu{m}$, t}he J and HK grisms were used in conjunction with the zJ and HK3 filters, providing bandpasses covering 0.94-1.51 and 1.25-2.34 microns, respectively. \edits{The utilized filters were necessary to limit the wavelength coverage to match those best suited to each of the gratings.} In this configuration the J/zJ combination provides a resolving power of 2400, with 2600 pixels/spectrum, and the HK/HK3 combination has a resolving power of 1400 with 800 pixels per spectrum \citep[]{mcleod_mmt_2012}. \edits{The slit aperture size (1.0$"$ by 0.603$"$) subtends 5.74$\times$10$^{6}$ cm by 3.46$\times$10$^{6}$ cm at 46P's geocentric distance of 0.0792 au for these observations.} For comparison, much previous work at similar wavelengths on other comets was completed with the SpeX instrument \citep{rayner_spex_2003} on the NASA IRTF, which has an average resolving power of $\sim$ 100 using the most-common 'prism' observing mode, or about a factor of $10-20$ courser resolution. There are key trade-offs though: while the MMT might have $\sim$4$\times$ the collecting area of the IRTF and is consequently able to observe at higher spectral resolution than SpeX on the IRTF, the IRTF is also at a much higher altitude allowing for a more reliable correction of atmospheric features, including observations at $\sim3.0\mu{m}$.

A summary of targeted spectral observations taken between 06:00 and 09:00 UTC is presented in Table \ref{tab:obs_log}. The observations of the science target were followed by observations of the stable high-quality Solar Analog star SAO 93936 (HD 28099, Hya 64, \edits{Bus, S.J.B., private communication}) nearby on the sky for proper correction of atmospheric absorption. \edits{This star and observation-reduction scheme has been used to great success on previous near-infrared observations of bright and faint solar system targets both \citep{noonan_search_2019,2019AJ....158..204S}}. (A normal 'bookend' observational plan was challenging to implement with the queue-based scheduling. The two different exposure times for the standard star observations at J-band are also a result of the queue-based scheduling after a shift change for the telescope operators.) The targets were dithered along the slit in an ABBA pattern for effective sky subtraction as well as to correct for any detector features. Observations of comet 46P/Wirtanen used a dither of $\sim$20.0" (the distance between the ``A" and ``B" positions), while the observations of the calibration star used a smaller $\sim$10.0" dither. We also obtained sky spectra, 1$^\circ$ north from comet 46P after every two observations, as an additional way to remove telluric features if needed. The spatial extent of the comet in the spectral data was such that with the 20" dither was sufficient to avoid significant contamination by doing an A-B (position ``A" minus position ``B") subtraction instead of an A-Sky (position ``A" minus an exposure taken far away on the sky) routine, especially so for a narrow nucleus-centered extraction as we implement below. At the crossover point between the two pointings (e.g. 10" from the pseudo-nucleus) the signal from the comet had dropped below $3-5\%$ of the peak signal, suggesting that contamination of the extraction area is essentially negligible. \edits{The spectral observations were obtained when 46P was near zenith at the parallactic angle, which was 90$\pm$3 degrees from the Sun/comet line, effectively perpendicular.}
 
Much of the reduction of both the comet and star data, including linearity corrections, flatfielding, and sky-subtraction utilized the pipeline described in \cite{chilingarian_data_2015}. While this pipeline can properly reduce, extract, and complete telluric corrections automatically on point sources, this was not possible for our science target. As a result, after allowing the pipeline to complete all steps mentioned above, the star and comet spectra were then traced and extracted using a series of custom-written scripts in Python. Each individual extracted comet spectrum (a combination of one 'A' and one 'B' frame) was then divided by the closest in time stellar observation, then all comet spectra at all wavelength settings (zJ and HK) were \edits{averaged} into a single master spectrum at a linearized resolution of $0.002\mu{m}/pix$.

As an auxilliary method to search for evidence of icy grains or grain break-down in the inner coma of 46P/Wirtanen, we obtained broadband images at J, H, and K-band with the MMT on the same night as our spectroscopic observations. These images were not photometrically calibrated as the primary goal was to search for morphological differences between them. The details of these observations are also listed in Table \ref{tab:obs_log}. The images were dark-corrected and a semi-randomized dither pattern was utilized to remove artifacts of the near-infrared detector. The J, H, and K-band images are shown in the top row of Figure 1. Using the full-width half-max of the field stars in each frame measured perpendicular to the direction of trailing, we find approximate seeing values in each filter of $FWHM_{J}\sim0.6-0.7\arcsec$, $FWHM_{H}\sim1.0-1.1\arcsec$, and $FWHM_{K}\sim0.7-0.8\arcsec$. Considering the short period between each set of exposures and their low airmass, it seems more likely that the H-band images are simply focused slightly less well than the J or K images rather than some very short term variation in seeing. (This is also suggested by the actual inner-coma profiles in the images themselves, but cometary comae are not gaussian in shape and thus cannot be used to measure seeing directly.) These seeing estimates strongly suggest that our small $0.603\arcsec$ extraction aperture used for our spectral data is well matched to the seeing in the near-infrared and thus really does capture the properties of the innermost coma.

\startlongtable
\begin{deluxetable*}{c|c|c|c|c|c|c}
\tablecaption{Observing log for the early morning of December 19, 2018. Seeing for the night was between 0.5 and 1.2" at optical wavelengths.}
    \label{tab:obs_log}
    \centering
    \startdata
        Target & Grism & Filter & Time (UTC) & Airmass & $N_{exp}$ & $\tau_{exp}$ (s)\\
        \hline
        SAO 93936 & J & zJ & 07:13 & 1.19-1.20 & 2, 2 & 9, 13 \\
              & HK & HK3 & 08:16 & 1.20-1.21 & 2 & 9 \\
        46P/Wirtanen & J & zJ & 06:23 & 1.00-1.08 & 12 & 120 \\ 
          & HK & HK3 & 07:03 & 1.03-1.07 & 10 & 120\\
         & -- & J & 04:59 & 1.01-1.01 & 10 & 15\\
         & -- & H & 05:11 & 1.00-1.01 & 20 & 15\\
         & -- & K & 05:30 & 1.00-1.00 & 20 & 15
    \enddata
\end{deluxetable*}

\subsection{Retrieved Spectra of Wirtanen's Inner Coma}
The extracted and calibrated relative reflectance spectrum of the innermost part of the coma of 46P/Wirtanen is shown in Figure 3. The retrieved spectrum is quite red (increasing reflectivity with increasing wavelength) and concave-down approaching neutrally-colored by H-band ($>1.95\mu{m}$). Beyond $2.2 \mu{m}$ there appears to be an increase in reflectivity, \edits{likely} thermal emission from the darker grains themselves which is discussed later or \edits{perhaps caused by poor calibration at longer wavelengths}. \edits{There is no obvious signature of the characteristic absorption features of water ice at 1.5$\mu{m}$ and 2.0$\mu{m}$, consistent with the NIR result of \citet{2021PSJ.....2..176P} obtained between December 13 and 17 2019 from the IRTF, \secondedits{6 days prior and 19 days after our observations}, at nearly identical observing geometry. } Setting an upper limit on the amount of ice content that could be in the observations we obtained of this inner most coma requires careful spectral modeling and subsequent interpretation, which is in Sections 5 and 6.


The obvious questions that \secondedits{stem} from this high-quality spectrum are: 1.) How much icy material could be hiding in this spectrum? 2.) What are the properties of the volatile-poor dust, and what does that imply for 46P/Wirtanen's source of water production and the processes acting within its innermost coma? To quantify the slopes of the spectrum over several wavelength ranges and thus \secondedits{quantify} the curvature of the spectrum, we adopt the $S_{\lambda}'$ framework of \citet{luu1990nucleus} over the ranges (1.17$\mu{m}$, 1.33$\mu{m}$), (1.49$\mu{m}$, 1.78$\mu{m}$), and (1.95$\mu{m}$, 2.20$\mu{m}$), broadly corresponding to the J, H, and K photometric bands (the latter until the $>\sim2.2\mu{m}$ increase in reflectivity). We retrieve $S_{J}'=5.9\pm0.3$, $S_{H}'=3.6\pm0.1$, and $S_{K*}'=-0.2\pm0.6$, with all reported values in $\%/(0.1\mu{m})$ and $1-\sigma$ errors. 
\secondedits{\citet{2021PSJ.....2..176P} measure their slopes through a linear fit to their data normalized at $1.0\mu{m}$ outside of areas of high telluric absorption and find \thirdedits{$S_{1.0\mu{m}}=2.1\pm0.3$ two days prior, on December 17, 2018}. If we replicate their methodology, we get $S_{1.0\mu{m}}=3.5\pm0.1$. Our reported error is strictly statistical while theirs incorporates variability from the multiple standard stars utilized. When coupled with the intrinsic slope variability in Wirtanen's spectrum seen in that paper (another $\pm0.3$ or so) and the spectral slope uncertainty in SpeX itself \citep{2020ApJS..247...73M} added in quadrature, we estimate an approximate $\sim0.6 / (0.1\%/\mu{m})$ in overall slope uncertainty comparing the two datasets. If this is a reasonable estimate, then the datasets are only in disagreement at the $\sim2-\sigma$ level. In other words, our dataset appears slightly redder than the reflectance spectrum of \citet{2021PSJ.....2..176P} at a level that is broadly within expectations.}

It is worth stating that the combination of the zJ and HK filter spectra combined easily and were consistent in slope at their overlap wavelengths prior to combination, which we view as good evidence that our calibration scheme produced reliable results. \edits{We also tried extracting the comet and star spectra over larger apertures and found} \secondedits{statistically identical spectral slopes}\edits{, but the low-altitude site of the MMT} \secondedits{combined with the finite dither distance along the slit} {made background subtraction and telluric correction of these larger aperture spectra rather challenging.} \secondedits{In other words, the spectra extracted from larger apertures were noisier near the features of interest but appeared to have the same spectral behavior (slopes, curvature) overall.} \edits{We estimate that each of these effects could be part of the difference between our spectrum and those of \citet{2021PSJ.....2..176P}.}

To quantify the approximate brightness of the comet throughout the near-infrared, we compared the brightness of the comet at each of the J, H, and K wavelength ranges to that of our standard star SAO 93936 observed with the same settings at similar airmass and closeby in time. For the innermost section of the coma of 46P/Wirtanen, we retrieve $m_{J}=10.85\pm0.03$, $m_{H}=10.28\pm0.05$, $m_{K}=9.99\pm0.04$, including the uncertainties on the data itself as well as the magnitudes of the standard star. These magnitudes are for the extraction area of the inner coma, which again subtends $0.603"$ in length with a $1.05"$ slit width. \edits{For a $\sim$600$m$ radius object with a traditional $p_V\sim0.04$ visible albedo, we would estimate that the nucleus of Wirtanen would have had $m_V \sim 13.31$, $m_J \sim 12.15$, and $m_K \sim 11.77$, assuming solar colors, indicating possible nucleus contributions of $20-30\%$ across our wavelength range. However, this is an upper limit considering that dust ``in front" of the nucleus from the perspective of the observer would block some of the light from it.}

\subsection{NIR Imaging Stacking, Processing, and Results}
After application of standard reduction for near-infrared images, we stacked all of the J, H, and K band images and averaged them. Alignment of the comet was attempted using multiple methods (aligning by brightest pixel within the inner coma, aligning by weighted flux, etc.) with similar results. The normalized and log-scaled images of comet 46P/Wirtanen are shown in the top row of Figure 1. The second row of Figure 1 shows the same \secondedits{images with an azimuthally averaged spatial profile subtracted from each} to highlight any asymmetries. The bottom row of that figure shows the ratios between each of the average images at each filter to highlight any potential gradients in retrieved grain colors that might indicate grain breakdown or changes in properties otherwise. \edits{Spatial profiles extracted from the bottom row of Figure 1 are shown separately in Figure 2.}

The anti-sunward dust tail is obvious in all of the images and is the only clear feature that remains after subtraction of the radial profiles. \edits{(The tail, jets, and similar features indicating an over-abundance of light would show up as positive (e.g. brightly colored) features above the dark background.)} The radial profile subtracted images and image ratios show artifacts within their respective seeing limits, inside of which interpretation of any structures is impossible. No obvious dust structures are shown that could be linked to rotation, either in these average composites or in inspection of individual frames (not shown), but this is not surprising due to the short temporal baseline probed by these observations. The image ratios (bottom row) are essentially all flat within the noise and after considering issues introduced by seeing.  (These differences in apparent seeing are the origins of the light/dark/ring features at the center of each image in this row.) Spatial profiles taken through these ratio images, \edits{shown in Figure 2}, and visual inspection give no clear indication for a change in grain reflective properties within the part of each image stack where the comet is well detected in both filters. \edits{The two best-focused filters, J and K, show a consistent-with-flat slope of their spatial profile cut across the near-nuclear region. A decreasing ice abundance from the nucleus would show increasing brightness in the H and K filters compared to J with nuclear distance, so this would argue against any large-scale transformation of dust properties within the studied region. However, the structure of the J/H and H/K ratios suggests that the H-filter images have a slower fall-off in brightness away from the Sun than the J or K filter images. It is hard to reconcile this structure as ice-related, as it would induce at least some features in the K-filter as well. Without more information available, we thus assume this is some artifact in the H-filter images related to their comparatively poor focus. The spatial profiles were all extracted over the same spatial extents shown in Figure 1, show little variation with small changes in angle, and are consistent with noise at their left and right edges. While our images detect the comet extremely well within several arcseconds of the nucleus (technically the optocenter), deeper imaging might have been able to push our sensitivity to significantly larger nucleus-centric distances -- perhaps a task for \textit{JWST}.}

\begin{figure}[ht!]
\plotone{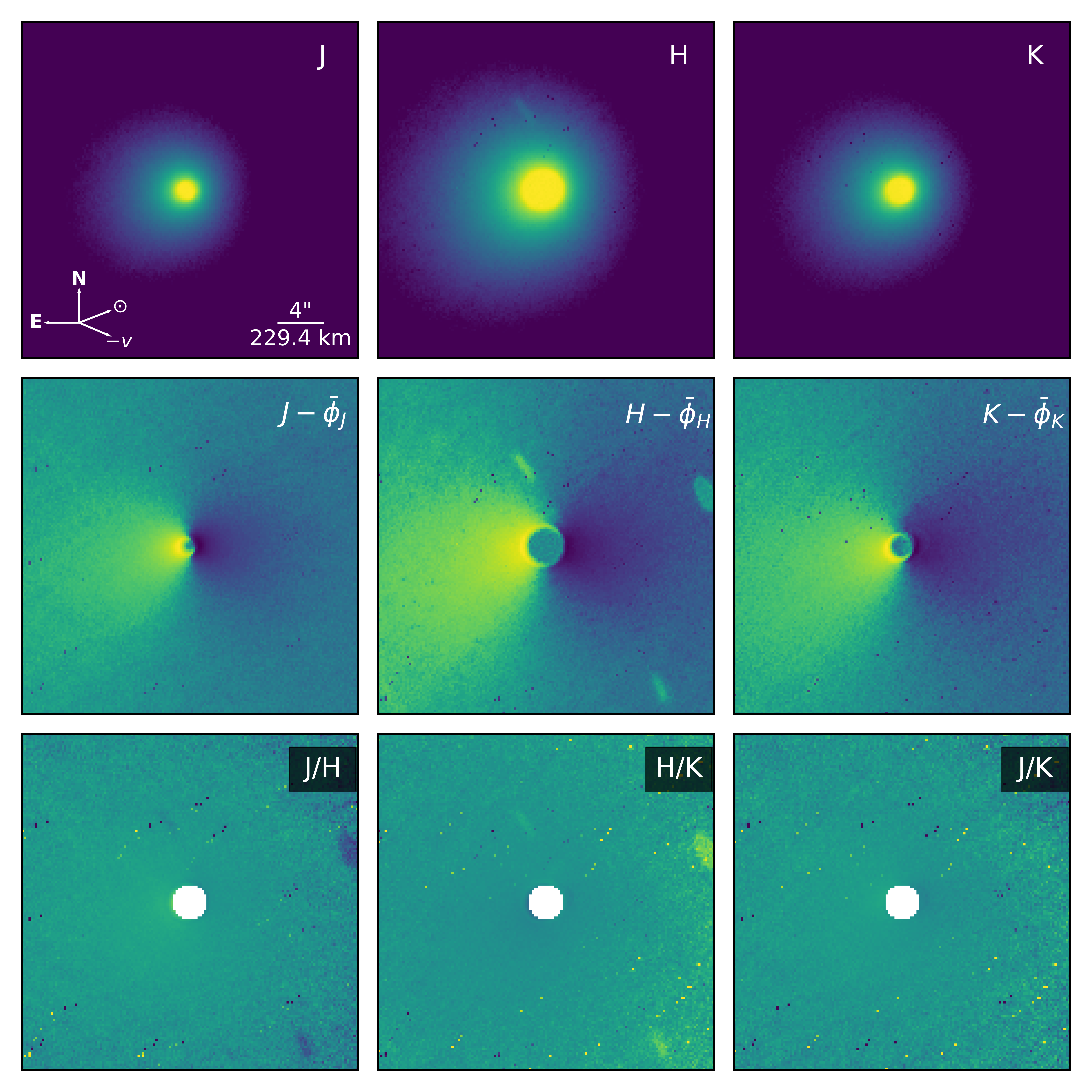}
\caption{Top Row: average images of 46P/Wirtanen normalized and shown on a logarithmic scale in the J, H, and K filters. \edits{The images are scaled identically in each row.} Middle Row: The same J, H, and K images shown after subtraction of an azimuthally averaged radial profile shown on a symmetric-logarithm scale. Bottom Row: The ratios between the images taken in each filter shown on a logarithmic scale \edits{between 0.3 and 3.0 times the median value in each image, as well as masked in the center, to enhance contrast.} \edits{The variations are indeed subtle and not a feature of the color map or scaling.} \thirdedits{Of particular note is the anti-sunward extension of the dust tail.} See text for more details and interpretation.}
\label{fig:JHK_images}
\end{figure}

\begin{figure}[ht!]
\plotone{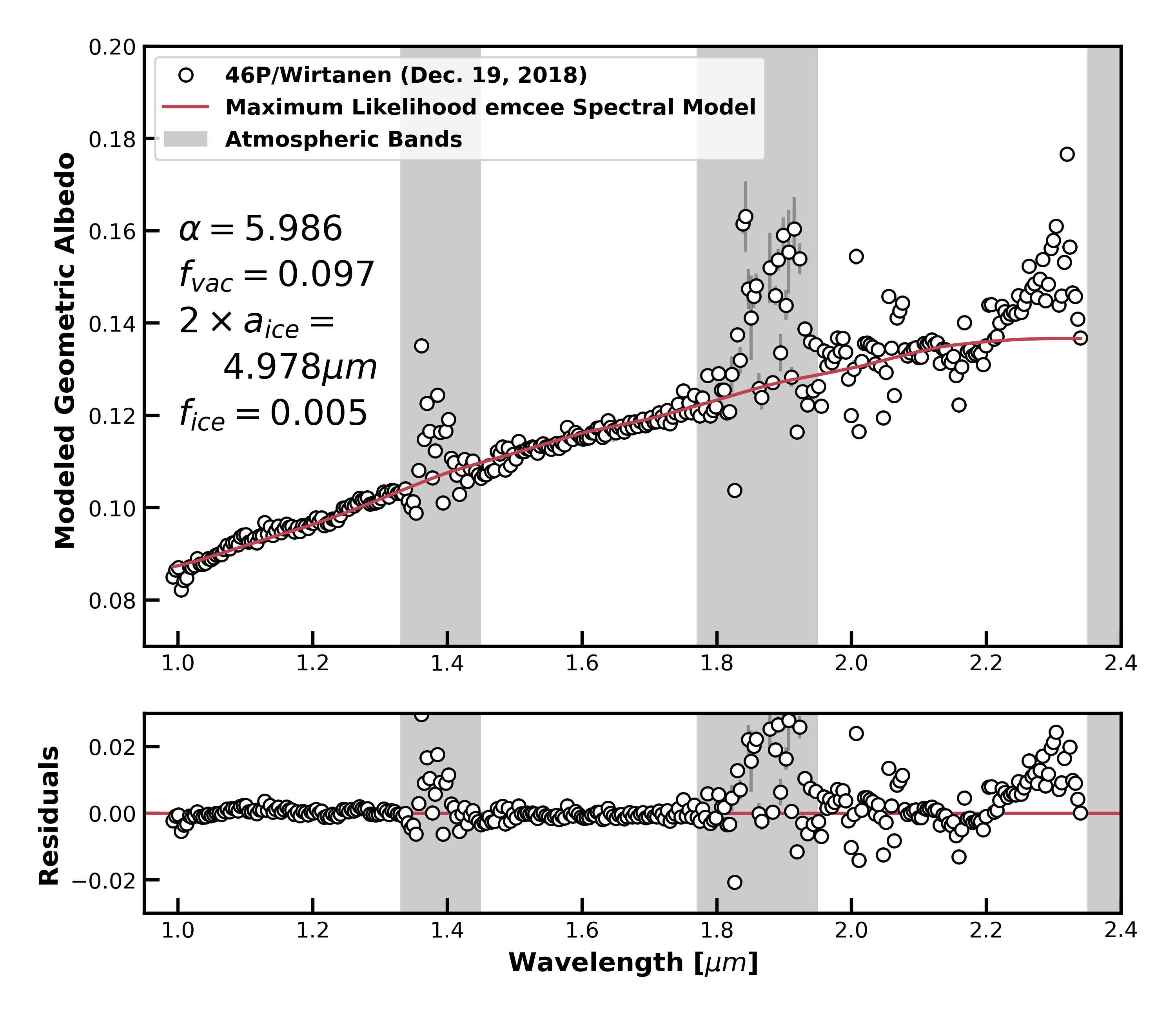}
\caption{Top: The near-infrared spectrum of 46P/Wirtanen's innermost coma is shown as black unfilled circles with error bars with the Maximum Likelihood spectral model from the MCMC package \textit{emcee} shown as a continuous curve in red. \edits{Bottom: The residuals (data $-$ model) are shown as a function of wavelength, with zero shown as a straight red line for reference.} The fit is shown to coincide with the data quite well, and as a result both the model and the data were moved out of normalized reflectance space by scaling both by the modeled geometric albedo at $\lambda=1.0\mu{m}$ of approximately $\sim0.087$. The high quality fit and the dark visible albedo together combine to suggest that our model is physically reasonable given the information we have available.}
\end{figure}

Water ice has absorption features at $1.5\mu{m}$ and $2.0\mu{m}$ in reflectance spectra, with an additional feature at $1.65\mu{m}$ if the ice is crystalline as would be expected at $\sim1.0$ AU. For the specific variants of the J, H, and K filters available for the MMIRS instrument, we would expect the presence of icy material to have no effect at J-band, moderate-to-strong effect at H-band, and strong effect at K-band. The lack of any clearly detectable gradient in grain color in our images of Wirtanen's inner coma as well as no obvious morphological differences between the filtered images themselves (after accounting for seeing \& focus) does not suggest any large changes in grain properties within $\sim15-20\arcsec$ of the nucleus. This indicates that any loss of icy material is subtle (\secondedits{if there isn't much ice to sublimate -- either through a low ice fraction in the individual grains or through a low overall abundance in the coma -- then the spectral changes that would result from its sublimation might be quite small}), the loss of icy grains happens \textit{very} close to the nucleus \secondedits{(short lifetimes for ``dirty" icy grains)}, or that \secondedits{the grain lifetimes are longer than the time it would take for the grains to leave our field of view}. However, such \secondedits{hypothetical long-lived} ice would still show spectral signatures \secondedits{if sufficiently abundant}, which we did not see clear evidence for in the previous sub-section and quantify in Section \ref{Modeling}. \edits{We note that \citet{2021PSJ.....2...92S} suggest that cometary hyperactivity falls along a spectrum such that each of these scenarios is \textit{a priori} plausible and allowed. Higher spatial resolution as might be facilitated by spacebased assets like \textit{JWST} could discriminate between these cases more directly.}

\section{Spectral Modeling}\label{Modeling}
\subsection{The Solid Fraction of Cometary Comae}
The reflected light from solid particles in the comae of active objects \secondedits{reveals} the compositions and size distributions of the individual particle populations. Of particular interest is separating the properties and abundances of the drier volatile-poor components from the icier volatile-rich components using near-infrared spectroscopy. 

The traditional way to model these spectra has been to select a featureless red-sloped refractory substance (almost always the amorphous carbon of \citealt{rouleau1991shape} or \citealt{edoh1983optical}) and a particular kind of water ice (e.g. the ``warm" ice of \citealt{warren2008optical} or the ``cold" ice, either amorphous or crystalline, from \citealt{mastrapa2008optical}), followed by selection of a scattering modeling methodology (Mie theory vs. \citealt{hapke2012theory}) and then a choice of mixing model (linear or intimate). While usually appropriate to try fitting the data with a suite of models with different physical understandings of what the dust \textit{could} look like, it is often the case that several of these models clearly fail to replicate the data, leaving only a handful to develop in more detail. For instance, if the spectrum is sufficiently curved (often becoming less red as a function of wavelength), then the refractory grains are likely at least partially in the Mie scattering size regime, making Hapke-style models hard to implement. If the characteristic water ice absorption features at 1.5 $\mu{m}$ and 2.0 $\mu{m}$ are not seen obviously above the noise, then the ice fraction cannot be that high considering the overall much higher albedo of ice compared to common cometary dust. That being said, within a particular class of model, the individual fit parameters are highly covariant with similar or complementary effects on the final output spectrum. A higher ice fraction could be hidden in the data if the icy grains were either sufficiently large\edits{-and-impure} or dominated by particles in the Mie regime that limit the depth of the water ice absorption features, while a low ice fraction leaves little ability to meaningfully constrain the properties of any icy grains that might be present. The size-frequency distribution and porosity of the refractory grains are similarly covariant as they both affect the overall slope and curvature of the spectrum. A proper understanding of these covariances and degeneracies and the interrelation between the modeling parameters is necessary to accurately assess what can be understood about the coma's properties.

\subsection{Applied MCMC Spectral Modeling Technique}
For the present study, we will employ \edits{the spectral modeling approach of} \citealt{protopapa_icy_2018} \edits{but with the fitting technique changed to use Maximum Likelihood Estimation (MLE) / Markov Chain Monte Carlo (MCMC) described below}. This is a natural choice as both the spectra of C/2013 US10 (Catalina) in that study and 46P/Wirtanen in our study are generally red-sloped with increasingly shallow and then neutral slopes at longer wavelengths, indicating to first order that a similar modeling procedure might be appropriate. Those authors assumed two major populations of grains: a power-law distribution of porous amorphous carbon grains (with the indices of refraction calculated assuming Bruggeman mixing, \citealt{bruggeman1935berechnung}) and a spatially separated population of single-sized small icy grains,  either pure or almost entirely pure, with both populations of grains having their reflectivities calculated using Mie theory. We can write the \textit{geometric albedo} of this mixture as:

\[p_{mixture} = (1 - f_{ice}) \times p_{dust} + f_{ice} \times p_{ice}\]

where $p_{dust}$ is the geometric albedo of the dust population, and $p_{ice}$ is the geometric albedo of the ice population. We can write the \textit{single-scattering albedos} of each population as:

\[w_{dust} = \frac{\int_{a_0}^{a_1} a^{2-\alpha} Q_{S}(a) da}{\int_{a_0}^{a_1} a^{2-\alpha} Q_{E}(a) da}\]

\[w_{ice} = Q_{S}(a) / Q_{E}(a)\]

where $Q_{S}(a)$ and $Q_{E}(a)$ are the Mie theory scattering and extinction efficiencies for a particle of radius $a$ given the appropriate optical constants. The factor $a^{2-\alpha}$ in the two integrals comes from the assumption that the particle population takes the form $n(a) = n_0 * a^{-\alpha}$, such that the cross section area of such a mixture would be $area = n_0 * a^{2-\alpha}$. (The factor $n_0$ is just a constant which appears on the numerator and denominator of the equation and is thus cancelled.) As in \citet{protopapa_icy_2018}, we take $a_0 = 0.5\mu{m}$ and $a_1 = 50\mu{m}$. To convert from single-scatteing albedos ($w$) to geometric albedos ($p$), we make the choice as made in previous studies to ignore phase angle effects of the incoming light, both for modeling simplicity and because our observations were near to opposition.

\[r_0 = \frac{1 - \sqrt{1 - w}}{1 + \sqrt{1 - w}}\]

\[p = \frac{1}{2}\times r_0 + \frac{1}{6}\times r_0^2\]

Those authors first fit their `ice-free' spectrum of C/2013 US10 (Catalina) taken at small heliocentric distances (interior to 2.3 AU) to determine the dust properties, and then used that dust model as an input into their ice-and-dust model for spectra taken beyond 3.9 AU where the ice is stable, as the surface temperature of the comet was below the sublimation point . Considering that our observations were only taken at one epoch, we have to simultaneously fit for the properties of the (dominant) dust and (trace) icy grains. Our fit parameters are thus the amorphous carbon  grain size-frequency distribution power law exponent $\alpha$, the volumetric porosity of those same grain $f_{vacuum}$, the size of the ice grains in question $a_{ice}$, and the areal fraction of the coma taken up by the \edits{pure} icy grains $f_{ice}$.

Previous studies have employed a least-squares fitting procedure to determine the best-fit properties of the grains in question. While this procedure is more than sufficient to answer certain scientific questions or if auxiliary information is available, a Markov-Chain Monte Carlo (MCMC) approach is likely more appropriate to constrain the range of possible grain properties given the rather hefty covariances between all the different fitting parameters described in the previous section. We implemented this using the Python-based MCMC package \textit{emcee} \citep{foreman2013emcee}. We first verified that our model could reproduce the fits of \citet{protopapa_icy_2018}. For our dataset, we assumed a \textit{uniform} prior distribution on each of the four fit parameters, utilized 1028 different walkers \edits{initialized around the best-fit least-squares spectral model}, and 7,500 iterations (which is $\sim 100 \times \tau_{autocorrelation}$ as recommended by the \textit{emcee} documentation). The maximum-likelihood spectral model and the ``corner" plot showing the distribution of model parameters are shown in Figures 3 and 4, respectively.

\begin{figure}[ht!]
\plotone{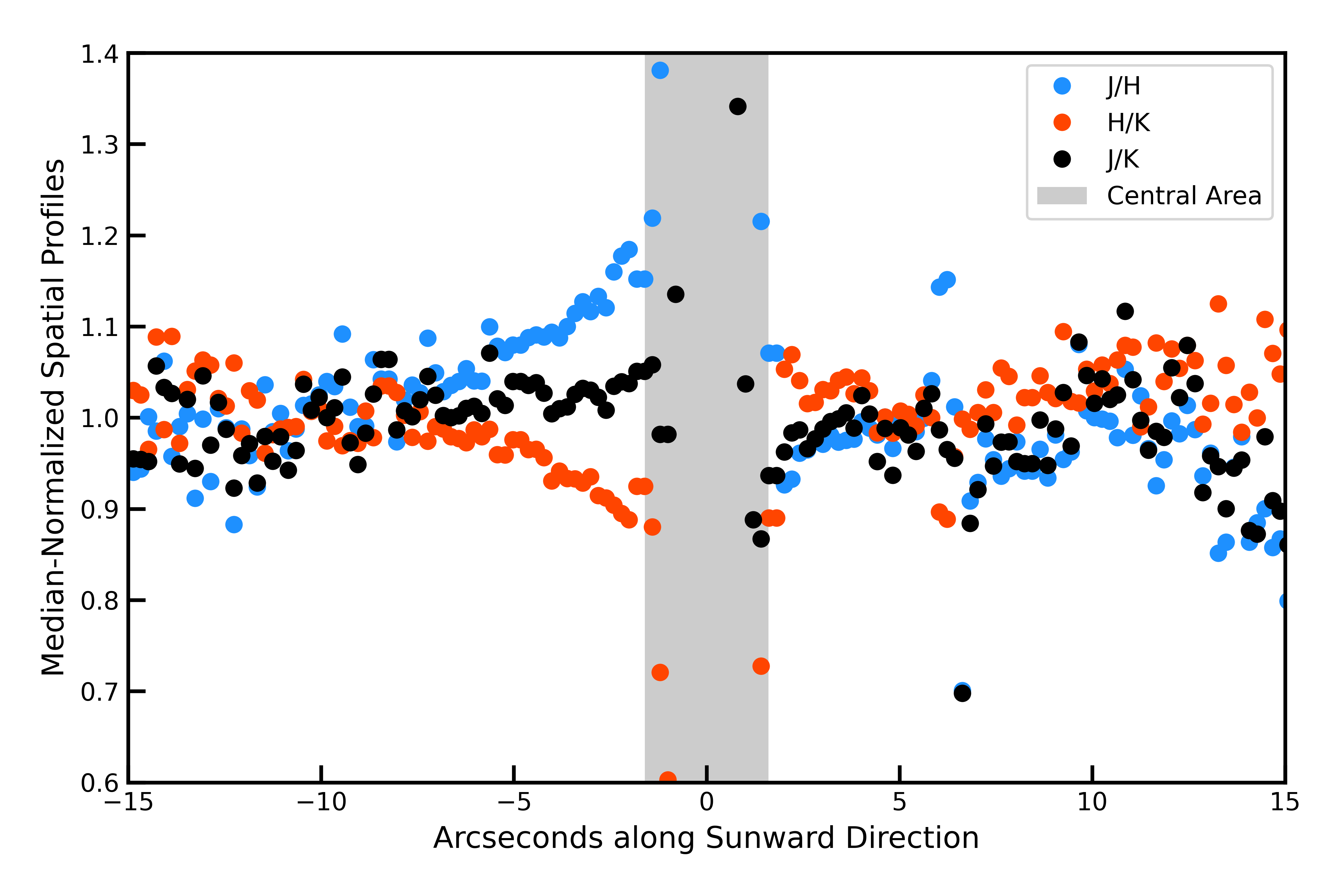}
\caption{\edits{Spatial profiles extracted along the Sun-Comet axis for the filter ratios shown in the bottom row of Figure 1. X-axis values increase towards the Sun and the Y-axis values were scaled by the median of each profile. Each profile is the average of three 1-pixel wide spatial profiles, including one which runs through the pixel corresponding to the brightest pixels in the original image and two profiles directly adjacent to it. Explanation and interpretation of the trends is in the text, but no \textit{unambiguous} signs for ice are apparent.}}
\label{fig:JHK_profiles}
\end{figure}

\begin{figure}[ht!]
\plotone{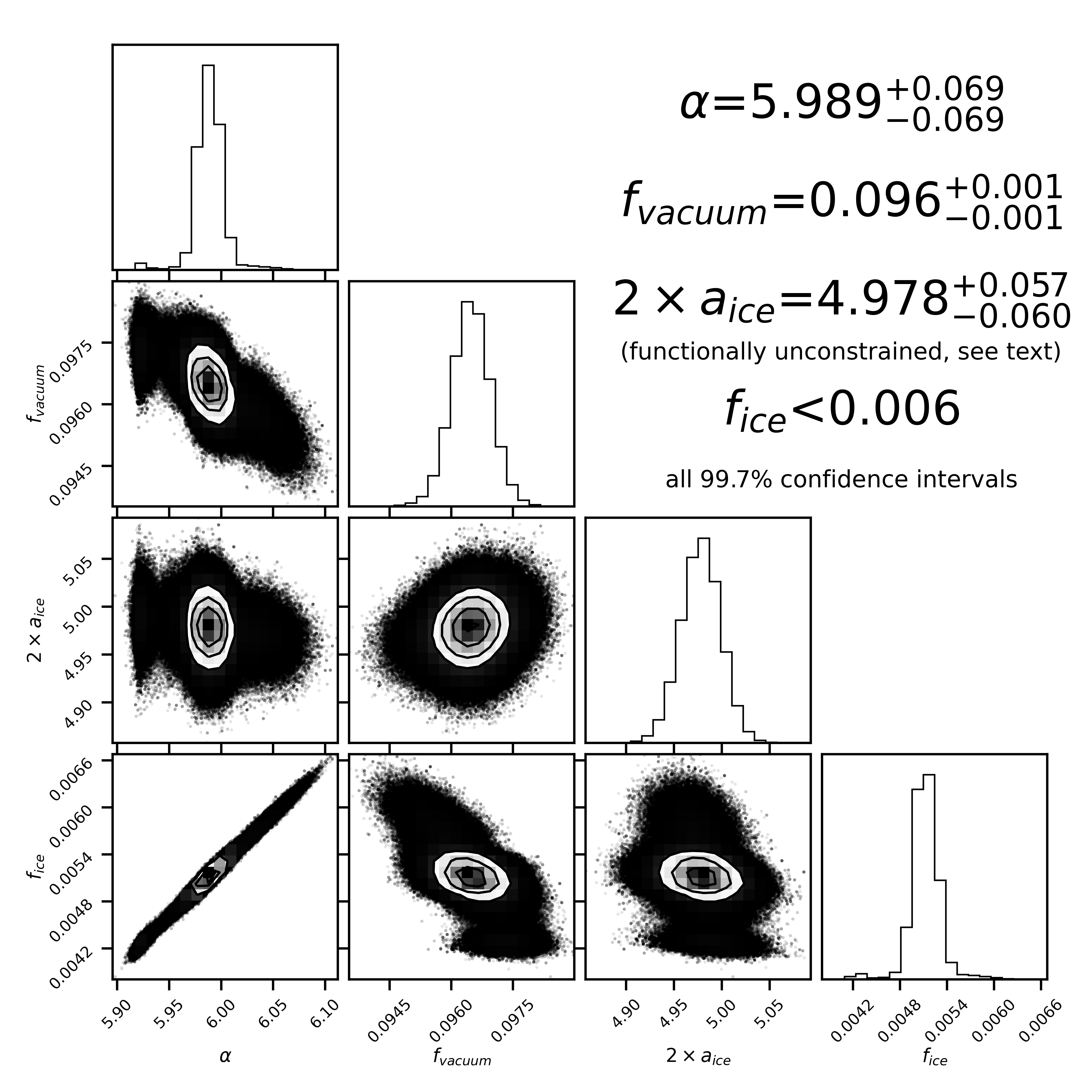}
\caption{The distribution of MCMC walker values for our four parameters are shown as a corner plot, with each of the two dimensional plots showing the inferred posterior distributions for each of the parameters against each other and at the top of each column is the histogram of marginalized values for each of our model parameters. As expected, the areal ice fraction is very low, less than $0.6\%$ at the $99.7\%$ confidence interval, and the ice grain size has a rather odd distribution due to the poor ability of the data to constrain that mode parameter. The dust grain size frequency exponent and volumetric vacuum fraction were constrained to $\alpha = 5.989_{-0.069}^{+0.069}$ and $f_{vac} = 0.096_{-0.001}^{+0.001}$ respectively, also at the $99.7\%$ confidence interval.}
\end{figure}

\subsection{Model Results and Inferred Maximum Ice Content}
As can be seen in Figures 3 and 4, the maximum-likelihood spectral model produces a great fit to the data short of $\sim2.25$ $\mu{m}$ and three of the four model parameters appear to be well constrained by the data. The dust grain size frequency exponent $\alpha$ (described formally in the previous section) was constrained to be $\alpha = 5.989_{-0.069}^{+0.069}$, and the volumetric vacuum fraction of those same dust grains was constrained to be $f_{vac} = 0.096_{-0.001}^{+0.001}$. At a 99.7$\%$ confidence level, the areal ice fraction of the inner coma of 46P/Wirtanen was less than 0.6$\%$. \edits{The least-squares-only fitting procedure that was used to initialize the walkers found similar values ($\alpha\sim5.96$, \thirdedits{$f_{vac}\sim0.097$}, $f_{ice}\sim0.01$ and $2\times a_{ice}\sim 5\mu{m}$). Due to the similarities of the best-fit and most-likely parameters, both the least-squares and maximum likelihood estimation produced reduced chi-squared values of $\chi^2 \sim 1.34$.} This is lower than the \secondedits{$1.4-2.2\%$} areal ice fraction reported by \citet{2021PSJ.....2..176P}, though we note that their modeling methodology differed somewhat from ours, \edits{including but not limited to a different method of implementing porosity as well as testing ice with $0.5\%$ volumetric contamination by dust in addition to the solely pure ice that our modeling has focused on}. The size of the dust, parameterized as $2\times a_{ice}$, might look as it if had converged to $2\times a_{ice} = 4.978_{-0.057}^{+0.060} \mu{m}$, but at such low areal ice fractions there is no apparent change in chi-squared or model spectrum when changing the parameters of the ice itself. \edits{(We remind the reader that very small ice grains (a few $\mu{m}$ or less) can rather diagnostically mute the depth of individual ice bands as shown in \citet{yang2014multi}, so the lack of constraints on ice properties from our fit is likely driven by a real lack of ice.)} The Maximum Likelihood values for the parameters are $\alpha = 5.986$, $f_{vac} = 0.097$, $2\times a_{ice} = 4.978\mu{m}$, and $f_{ice} = 0.005$. (The Maximum Likelihood values do not have to sit at the exact median of the parameter distributions, but in this case they're quite close.) As expected, the fit parameters are highly covariant (e.g. they deviate from smooth 2-D gaussians in Figure 4), suggesting that the more sophisticated \edits{fitting approach} was appropriate \edits{(and might be even more so in cases with clear ice detections)}. Perhaps most interestingly, a steeper dust size distribution (higher $\alpha$) could accommodate (hide) more ice, which might be of use to future works attempting to constrain these parameters simultaneously. It is worth stating explicitly that choices of different prior distributions, a different number of walkers or iterations, or different initial conditions did not affect the retrieved spectral modeling results.

One immediate question is why the areal ice fraction did not converge to 0.0$\%$ as opposed to being clustered slightly above 0.0$\%$. Comparison of models with the best-fit dust parameters and 0.0$\%$ ice to those with the best-fit dust parameters and 0.6$\%$ ice only vary in slope at a $\sim1\sigma$ level. Considering the brightness of the target and the number of spectra combined to make our master spectrum, even the difference in spectral slope of the Sun versus that of our solar analog SAO 93936 becomes apparent -- not even to mention the subtle slope changes that can result from imperfect telluric correction of a reflectance spectrum \textit{away} from the telluric bands. \edits{In other words, with no evidence for water ice in our data, even slight imperfections in telluric or solar analog corrections need to be considered more carefully than usual in interpretation of modeling results.}
 
\section{Results}\label{Results}
\subsection{Spectral Properties, Ice Upper Limits, and Observability Considerations}
As described in the previous section, our spectral modeling essentially confirmed what a preliminary inspection would suggest: there is no obvious evidence for water ice in our near-infrared spectrum of 46P/Wirtanen. \edits{This is in agreement with the conclusions of \citet{2021PSJ.....2..176P}.}
Observations (in reflected light, at least) are biased \textit{in favor} of the detection bright-and-blue pure-ice water grains, so the fact that our spectrum both does not show any absorption features that could be attributed to water ice and that it is as red as it is leaves few options. If a possible population of water ice grains is present, we constrain its areal fraction to be less than 0.6$\%$. If water ice was mixed in with impurities, it would warm and sublimate even faster, and might not even make it far enough from the comet to leave our slit (using the sublimation model of \citealt{protopapa_icy_2018}), and that is setting aside the fact that previous observations have suggested that nearly-or-fully pure icy grains are expected based on previous positive detections remotely and in-situ \edits{for other comets}. \edits{Like \citet{2021PSJ.....2..176P}, we are essentially better able to rule out pure water ice grains than impure ones for detectability and lifetime reasons.}
In summary, our retrieved reflectance spectrum appears to be a fairly normal spectrum for a cometary comae comprised of small dusty grains with no ice.

Our spectra were extracted in a rather narrow 1$\arcsec \times$0.6$\arcsec$ window centered on the photocenter -- the brightest part of the coma which is commonly presumed to be the approximate location of the nucleus. While in principle this is a good assumption, radiation pressure will move the optocenter slightly tailward due to radiation pressure upon the dust grains. Furthermore, while our observations are of the innermost part of 46P/Wirtanen's coma, it is also worth stating explicitly that the dust in between the observer and the photocenter contributes, such that drier outer coma material could be muting the signatures of a hypothetically icier inner environment. Those two particular issues are present in all attempted observations of the inner nuclei of comets, but this particular scenario also has the benefits of Wirtanen's extreme close approach and our observations taking place in the near-infrared. For a given coma brightness or dust production, moving the comet closer to the Earth (e.g. decreasing its geocentric distance) will make that same coma subtend a larger area on the sky, mitigating the issues that might cause outer coma contamination of our inner coma centered spectrum. \edits{(The nuclear signal would \textit{not} be muted in the same way, which is why our \textit{possible} nuclear contamination is higher than that of \citet{2021PSJ.....2..176P} when combined with their larger extraction aperture.)} Furthermore, while Wirtanen was still quite obviously extended in our NIR data,  dust at typical cometary sizes ($0.5-1.0\mu{m}$) will be less important at wavelengths larger than its size ($>1.0\mu{m}$) than comparable to it. A full accounting of all these various effects is beyond the scope of this paper (it likely would require a full 3D radiative transfer model of the coma along and near the line of sight), but the fact that the along-slit profiles of Wirtanen in our data still show such an obvious central condensation would argue against a large contribution from outlying dust which would by necessity be broader in spatial distribution on the sky. We thus conclude that our observations do probe the innermost coma with some undetermined, but small, amount of contamination.

\subsection{Ice Production Rate Estimates}

With our observed J, H, and K magnitudes we can derive the cometary dust production proxy Af$\rho$ \citep{a1984comet} for the night of December 19, 2018. We must implement the rectangular slit correction described in \citealt{am1984comet}, as our magnitudes are derived from an integrated portion of the slit and not circular aperture photometry. This process requires finding the filling factor of the observations and deriving the corresponding circular aperture radius with the same filling factor, using equations 1,2, and 3 of \citealt{am1984comet}. Without this correction it would be difficult to compare our Af$\rho$ values to those derived from images.

For these observations our aperture size subtends 1.0" in width by 0.603" in height, or 5.74$\times$10$^{6}$ cm by 3.46$\times$10$^{6}$ cm at 46P's geocentric distance of 0.0792 au.  Using the correction factor from \citealt{am1984comet} we find that the corresponding $\rho$ value is \edits{2.57$\times$10$^{6}$ cm}. With this $\rho$ we can then proceed to calculate Af$\rho$ using our derived J,H, and K magnitudes as well as the J, H, and K apparent magnitudes for the Sun from \citealt{willmer2018absolute}. For our J,H, K magnitudes of 10.85, 10.28, and 9.99 we find Af$\rho$ values of \edits{784, 1166, and 380 cm, respectively}. 

By using the geometric albedo from the MCMC model we can then derive dust production rates, and in turn the upper limit on the icy grain production rate. The dust production rate can be written as a function of Af$\rho$:
\begin{equation}
    Q_{dust} = \frac{2Af\rho r_{grain}v_{dust}\rho_{grain}}{3q_{\nu}}
\end{equation}
as detailed in \citealt{lorin2007search}, where $\rho$ is the aperture size, r$_{grain}$ is the grain radius, v$_{dust}$ is the dust velocity in m s${-1}$, $\rho_{grain}$ is the grain density, and q$_{\nu}$ is the geometric albedo. We implement a grain radius of 0.5 microns, our previously established $\rho$ of \edits{2.57$\times$10$^{6}$ cm}, a grain density of 1000 kg m$^{-3}$ from \citealt{meech1997large}, and our best fit geometric albedo 0.08. However, the dust velocity is less certain. Previously in the comet literature dust velocities of 465r$_{H}^{-1/2}$ ms$^{-1}$ \citep{delsemme1982chemical,lorin2007search} have been implemented, but the 2018 apparition of 46P has previously been shown to have dust velocities that are much lower, in the tens of m s$^{-1}$ \citep{farnham_first_2019} to 162$\pm$15 s$^{-1}$ \citep{farnham2021narrowband}. For the \edits{expected quiescent} velocities on the order of 30 m s$^{-1}$ we find dust production rates between \edits{46-142 kg/s}, while for an outburst-like velocity of 162 m s$^{-1}$ we find dust production rates of \edits{250-765 kg/s}, with the H band magnitude producing the highest derived rate. These dust production rates are on the order of what has been previously calculated for 46P \citep{farnham1998narrowband,fink1998spectroscopy,kidger2004dust} \edits{with the quiescent activity velocity representing a better match.}

With these dust production rates determined, we can then place a conservative upper limit on the mass of \textit{pure or nearly-pure} icy grains that are present in the near-nucleus coma of 46P/Wirtanen. With the MCMC derived upper limit on the areal fraction of icy grains at $\leq$0.6\% that of the dust, and assuming that the icy grain size should be similar to the dust grain size and a similar density \secondedits{($\rho$ = 1000 kg m$^{-3}$) i.e. pure ice grains with no porosity)}, we then determine that a conservative upper limit for the icy grain production rate is \edits{$\leq$ 4.6 kg s$^{-1}$}, using the highest dust production rate derived from the H band magnitude. Even if we assume that these grains are pure water ice and will sublimate within the field of view of the SWAN Lyman-$\alpha$ measurements from \citealt{combi_comet_2020} the icy grain contribution to the total water production rate is \edits{$\leq$ 1.5$\times$10$^{26}$ mol s$^{-1}$,  barely 1 percent of the measured water production rate at perihelion for 46P}. \edits{We note explicitly that less pure ice-bearing grains could contribute water production as well, should they exist in a significant quantity, but we are unable to constrain their abundance well.} This tight limit on the contribution to the water production rate from an extended source of \edits{pure or nearly-pure ice} grains has implications for calculations of the active area of the nucleus and 46P/Wirtanen's evolution as comet in recent apparitions, which we discuss in the next section. 

\section{Discussion}\label{Disc}
The lack of any distinctive icy grains in the near-nucleus coma of 46P/Wirtanen during the 2018-2019 apparition illustrates a new chapter in the comet's activity. With small ($\sim$ micron sized) \secondedits{pure} icy grains eliminated as a significant source of water in the bulk coma we will briefly theorize several potential implications for the nucleus properties of 46P. 
 
\begin{figure}
\centering
    \plotone{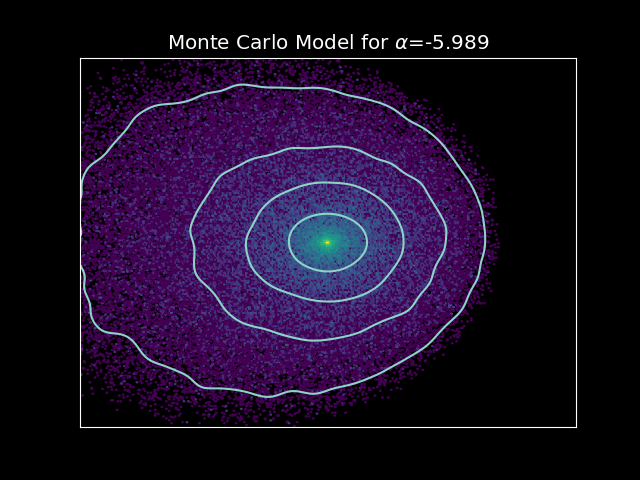}
    \caption{Monte Carlo icy grain model for grain sizes randomly sampled from the size frequency distribution governed by the $\alpha$=-5.989 power law. All particles are subject to gravitational force from the comet nucleus, pressure from the expanding coma gas (assumed to be \ce{H2O}), and solar radiation pressure at 1 au. \secondedits{The }plot is 15000 km by 15000 km and shown in a logarithmic scale to match the top row of images shown in Figure \ref{fig:JHK_images}. The model results have been rotated to match the Earth viewing geometry at the time of observation, a phase angle of 18 degrees and a orbital difference angle of -16 degrees. \thirdedits{The general morphology of the model is in agreement with observations shown in in Figure \ref{fig:JHK_images}.} }
    \label{fig:monte_carlo_model_results}
\end{figure}

The first and most obvious of the implications of a lack of small water ice grains to provide a distributed source is that the entire water production rate of 46P/Wirtanen must be attributed to the surface of the nucleus. Even if we implement the maximum water production rate for 46P during the 2018-2019 apparition measured by \citealt{combi_comet_2020} of 1.6$\times$10$^{28}$ measured on December 22, 2018, a radius of 630 meters \footnote{\footurl}, and a unit production rate of 5$\times$10$^{21}$ \secondedits{m$^{-2}$ s$^{-1}$} of water ice sublimation from \citealt{cowan1979vaporization} at 1 au for a low albedo nucleus we find that a maximum active area of $\sim$3.2 km$^{2}$ is required, approximately 64\% of 46P's surface area when a spherical nucleus is assumed. We note that our non-detection of icy grains directly conflicts with \citealt{knight2021narrowband}'s interpretation of the tailward OH distribution they observed, which the authors had attributed to icy grains pushed tailward by solar radiation pressure. This leads us to question if the \secondedits{ice-bearing grains} that are responsible for the OH distribution are too large to be detected in our spectrum. \secondedits{\citet{2021PSJ.....2..176P} suggested that contaminated icy grains with more than $0.5\%$ dust-by-volume might explain Wirtanen's hyperactivity without detectable spectral features, so if those were extant, they might be compatible with the tailward OH distribution should they be of the right size to move the correct distance from the nucleus before sublimation.} \secondedits{While large very icy particles should still produce detectable absorption features, larger `dirty' grains could significantly contribute to ongoing OH production while being too dusty to produce clear spectral features.} To determine which sizes of icy grains can be pushed substantially tailward in the inner coma, less than 1000 kilometers from the nucleus as shown in Figure 3 of \citealt{knight2021narrowband}, we performed Monte Carlo modeling of a range of icy grain sizes, from 1 micron up to 10 cm, ejected from the nucleus at random orientations and times and accelerated by sublimating gas. The model calculates the radiation and gas pressure on each particle at 1 second intervals, with an adaptive timestep that increases with decreasing gravity and sublimation pressure. A snapshot of particles is obtained at a random period between 0 and 10$^{5}$ seconds, the time to reach 10,000 km assuming a 100 m/s velocity. It is worth noting that the maximum velocities reached by the micron-sized grains are between 100 and 140 m/s.  Of the 6 orders of magnitude in size that we sample only the grains smaller than 100 microns matches the distribution shown in \citealt{knight2021narrowband}; larger sizes remain more tightly distributed relative to the nucleus, with little acceleration from gas drag to push them farther from the nucleus. Further, we show in Figure \ref{fig:monte_carlo_model_results} that when plotting the spatial distribution of dust particles randomly assigned radii from the size frequency distribution $\alpha$=-5.89 the extent of the coma is consistent with that shown in Figure \ref{fig:JHK_images}. 
 
Second, and perhaps the most supported by additional observations of volatiles and hypervolatiles in the 2018-2019 apparition, is that the hypervolatile reservoirs typically responsible for lifting icy grains from cometary surfaces were depleted and/or \edits{inaccessible}. The lack of significant CO \citep{noonan2021fuv,mckay2021quantifying,biver2021molecular}, which has been suggested as a substantial contributor to lifting icy grains at larger heliocentric distances \citep{kelley2013distribution, ahearn_comets_2011, protopapa2014water} would support this conclusion, \secondedits{but the inferred detection of CO$_2$ \citep{bauer2021neowise} complicates this.} This would not necessarily exclude a second option, where the number of icy grains capable of being lifted are themselves depleted. Given that the Af$\rho$ and dust production rates that we measure are consistent with both contemporaneous and past observations \citep{knight2021narrowband,stern1998hst,farnham1998narrowband,kidger2004dust}, it is difficult to reconcile either of these as being a dominant factor in decreasing the abundance of icy grains in the coma and driving the change in water production rate of 46P over the last 25 years. One possibility is that the series of post-perihelion outbursts observed during the 2002 apparition were global outbursts \citep{combi_comet_2020,kidger2004dust}, shedding large amounts of icy grains and depleting large swaths of the nucleus of the volatiles typically required to lift those that were left. Larger grains lofted by the heavy activity would lose their ice before re-accreting on the comet, mantling the surface and decreasing production rates. \edits{(We refer the reader again to \citet{2021PSJ.....2...92S} for a discussion of different scenarios and implications of icy grain production for context.)} However, our observations can only provide constraints on the smallest icy grains in the coma. One important caveat to make is that our \secondedits{imaging} observations are sensitive to small icy grains, but that there may still be a significant population of $\sim$mm-cm sized grains. As stated previously, the fact that the tail of Wirtanen is starting to form in our images suggests that the light in them is likely dominated by smaller ($10-100\mu{m}$ or so) grains, in addition to the fact that observations in the NIR are simply most sensitive to NIR-sized grains as general rule of thumb. These grains would have longer lifetimes than the micron-sized grains the MMT \secondedits{imaging} observations are sensitive to and ultimately would have little effect (\edits{$<$}10\%) on the bulk water production rate unless they were abundant i.e. N$_{ice}$ $\sim$10$^{11}$ grains in the inner coma requiring a production rate of Q$_{ice}$ $\sim$17 kg s$^{-1}$). 

Detections, or the lackthereof, of icy grains are highly useful to constrain the bulk properties of the material lofted from the nucleus and to understand the global volatile budget of comets (and Centaurs), but circumstances often prevent clear determination of their existence or properties. While this apparition of 46P/Wirtanen was undoubtedly an excellent opportunity to probe its inner coma, the low geocentric distance and thus heliocentric distance made any icy grains much shorter lived and thus more challenging to characterize. Simultaneously, observations at larger heliocentric distances of Centaurs or Long Period Comets might allow a more sizable population of icy grains to build up, but clarity over their spatial distribution (e.g. when they were released from the nucleus, how uniformly, etc.) is lost in addition to the targets themselves inherently being dimmer. There is simply limited data available and only a handful of positive clear detections of icy coma material to guide future observations to be more useful. Studies of other long-period comets or hyperactive JFCs that cross the boundary where ice can sublimate vigorously could be quite useful in understanding where and when are the most likely places to search for these grains.

Another issue is the challenges with detecting \textit{larger} and intimately mixed icy grains. Small and pure icy grains can have relatively large impacts on the reflectance spectrum of a coma even with small abundances, but mixing ice into a darker material (e.g. amorphous carbon) and placing it within larger grains (that may be less prominent at common observing wavelengths, \edits{e.g. millimters and larger}) both can act to make extant ice much more challenging to detect. If such grains do exist (perhaps as suggested by the radar-discovered ``skirt" of larger grains around Wirtanen\footnote{ see \url{https://wirtanen.astro.umd.edu/46P/46P_status.shtml}, post on December 21, 2018}), then they might be able to supply some extra volatile production. As mentioned multiple times previously, the less pure the ice content of these lofted materials, the darker the albedo and thus the shorter the lifetime against sublimation. In other words, while our current observational schemes are likely not well suited to directly characterizing these hypothetical large intimately mixed grains, our current estimation of their properties would not make them very stable. \citet{2021PSJ.....2..176P} come to the same conclusion based on their observations and modeling: small \edits{pure} ice grains have to be very, very uncommon, but larger particles \secondedits{or sufficiently contaminated smaller particles} could supply the extra water production. At time of writing, the general understanding is that these icy grains in the comae of comets should be small and relatively pure, so the possible existence of a larger icy grain population either implies that Wirtanen is an outlier or that many comets have an as of yet undetected large ice-bearing grain population.

The work of \citet{sunshine2007distribution} at Tempel 1 and subsequent works suggested that icy grains of a $\sim\mu{m}$ in size are a common component for many cometary nuclei either at depth (e.g. Tempel 1, \citet{sunshine2007distribution}) or prior to significant thermal processing of the upper layers (e.g. C/2013 US10 (Catalina), \citet{protopapa_icy_2018}.) If we suppose that Wirtanen previously could produce a large icy grain population but can no longer, perhaps the lofted materials that form the solid coma are now being sourced from a shallower depth. If true, that could provide a natural explanation for the decreasing production rates of the comet epoch-to-epoch \citep{knight2021narrowband}. Outbursts or localized areas of activity would by necessity probe different depths than the bulk activity, which our particular very-inner-coma observations would likely not be very sensitive to unless they had happened so recently that the ejected material had not left the field of view. If this is the general scenario, then outbursts might be able to display evidence for icy grains that quiescent or stable activity might not, which does appear to be the case at 17P/Holmes \citep{yang2009comet}, but it is unclear if 17P's massive outburst is driven by the same processes.

We note that it is tempting to attribute the decline from hyperactivity of 46P/Wirtanen in the last few apparitions to decreased abundances of hypervolatiles capable of lifting significant amounts of icy grains, but to definitively prove such a connection requires both an active detection of the hypervolatiles (CO$_{2}$, CO) and of icy grain absorption features at a range of heliocentric distances and, ideally, multiple orbits. Such an observing campaign would cover decades and require both NIR and IR sensitivity, but offers an opportunity to confirm icy grains as a symptom of hypervolatile activity that promotes increased water production. The launch \edits{and successful Cycle 1} of the James Webb Space Telescope (JWST) in \edits{the} last year should provide an excellent platform to carry out this set of observations free from telluric interference, with the NIRSpec instrument sensitive to the strong 3 $\mu$m water absorption band and the Mid-Infrared Instrument capable of detecting CO and CO$_{2}$ in addition to the 6 $\mu$m water absorption band JWST is well suited to probing cometary icy grain trends.

\section{Summary}\label{Summary}
Near infrared images (JHK) and spectra ($1.0-2.3\mu{m}$) of comet 46P/Wirtanen were obtained using the MMIRS instrument on the MMT during the comet's historical close approach to the Earth on December 19, 2018. The reflectance spectra of the inner coma were fit using a Markov Chain Monte Carlo model that samples particle size distribution, albedos, and relative fractions of carbon and ice. From our observations and other considerations we derive the following conclusions:
\begin{enumerate}
    \item The NIR inner coma of 46P is featureless, red, and concave-downwards with slopes of $S_{J}'=5.9\pm0.3$, $S_{H}'=3.6\pm0.1$, and $S_{K*}'=-0.2\pm0.6$ in units of \%/($0.1\mu{m}$) in the notation of \cite{luu1990nucleus}. No signatures of ice are obvious in the data.
    \item Images in the J, H, and K filters and comparisons between them find no evidence for changes in morphology at different wavelengths or changes in bulk grain properties within the field of view, arguing against any (large) inner coma grain processing from grain breakdown or ice sublimation.
    \item MCMC modeling of the retrieved reflectance spectra find that the inner coma is dominated by porous refractory grains. The dust population has a size frequency exponent $\alpha$ of $5.989_{-0.069}^{+0.069}$\edits{, a volumetric vacuum fraction $f_{vac}$ of $0.096_{-0.001}^{+0.001}$,} and a geometric albedo of $\sim0.08$ at $1.0\mu{m}$. We place a 99.7\% confidence upper limit of $<0.6\%$ on the areal fraction of \secondedits{pure} icy grains, and are as a result unable to constrain the properties of any icy grains that do exist. Uncertainties are largely dominated by systematic effects due to the high brightness of the target.
    \item Our calculated inner-coma magnitudes of $m_{J}=10.85\pm0.03$, $m_{H}=10.28\pm0.05$, $m_{K}=9.99\pm0.04$ derived from our spectral observations produce slit-corrected Af$\rho$ values of  \edits{764, 1166, and 380 cm}, respectively. 
    \item Using the Af$\rho$ and best fit geometric albedo we find dust production rates between \edits{46-142 kg s$^{-1}$} for the expected quiescent dust velocities of 30 m $s^{-1}$, from \cite{farnham_first_2019}, and \edits{250-765 kg s$^{-1}$} for the observed outburst velocity from \cite{farnham2021narrowband} of 162.5 m $s^{-1}$.
    \item We find that the upper limit for \secondedits{pure} water ice grain production is \edits{$\leq$4.6 kg s$^{-1}$}, and is therefore not a significant source of water for the global water production rate, requiring that 64\% of 46P's surface is actively sublimating water vapor. 
\end{enumerate}

The existence and properties of icy grains in cometary comae represent a key attribute of some comets, though it is not yet entirely clear if that characteristic is more indicative of a primordial structure or acquired, and ultimately lost, via thermal evolution of the nucleus. Our study represents one small set of observations obtained during an excellent, but singular, apparition, of just one comet. Fully addressing the icy grain dilemma of comets will require similar NIR observations of both JFC and isotropic comets to search for patterns in orbital evolution and icy grain abundance. Ground-based resources like NASA's Infrared Telescope Facility, the MMT, the Large Binocular Telescope, and space-based resources like JWST provide ample opportunities to sample these comet populations for years to come. Instruments that provide low-resolution spectroscopy or imaging near $\sim3.0\mu{m}$ might allow for more constraints on more objects (the water ice absorption at $\sim3.0\mu{m}$ is a factor of $\sim100\times$ stronger than those at $1.5\mu{m}$ and $2.0\mu{m})$, but this is a current area of limited availability. JWST is especially critical to provide sensitivity to detecting water ice grains that are a factor of ten larger than those probed by J, H, and K spectra.

\textbf{Acknowledgements}: Observations reported here were obtained at the MMT Observatory, a joint facility of the University of Arizona and the Smithsonian Institution. We would also like to thank Brian Jackson for help in use of the \textit{emcee} package \citep{foreman2013emcee}, and Cassandra LeJoly and Nalin Samarasinha for helpful discussions regarding the observations and our interpretations of them.




\bibliographystyle{aasjournal}


\end{document}